\documentclass[10pt,journal, twoside]{IEEEtran}
\usepackage{cite}
\usepackage{hyperref}
\usepackage{amssymb}
\usepackage{amsmath}
\usepackage{graphicx}
\usepackage{graphicx,subfigure}
\usepackage{amsmath,amssymb}
\usepackage{cite}
\usepackage{url}
\usepackage{placeins}
\usepackage{xr}
\usepackage[table]{xcolor}
\usepackage[T1]{fontenc}
\usepackage[utf8]{inputenc}
\usepackage{tabularx,ragged2e,booktabs,caption}
\usepackage{array,booktabs,arydshln,xcolor}
\newcolumntype{C}[1]{>{\Centering}m{#1}}

\usepackage{slashbox}
\usepackage{adjustbox,lipsum}

\usepackage{cite}

{}
\usepackage{suffix}
\newcommand\MeijerG[8][]{%
  \mathsf{G}^{#2,#3}_{#4,#5}\MeijerM[#1]{#6}{#7}{#8}}

\WithSuffix\newcommand\MeijerG*[7]{%
  \mathsf{G}^{#1,#2}_{#3,#4}\MeijerM*{#5}{#6}{#7}}
	
\WithSuffix\newcommand\MeijerE*[7]{%
  \mathsf{E}\MeijerM*{#5}{#6}{#7}}

\begin{document}

\title{An Image Encryption Algorithm Based on Chaotic Maps and Discrete Linear Chirp Transform}

\author{Osama A S Alkishriwo \thanks{Osama A S Alkishriwo is a lecturer with the Department of Electrical and Electronic Engineering, University of Tripoli, Libya, Email: alkishriewo@yahoo.com}$^1$}

\maketitle
\markboth{Almadar Journal for Communications, Information Technology, and Applications}%
{Almadar Journal for Communications, Information Technology, and Applications}

%\IEEEtitleabstractindextext{
\begin{abstract}
%\boldmath
In this paper, a novel image encryption algorithm, which involves a chaotic block image scrambling followed by a two--dimensional (2D) discrete linear chirp transform, is proposed. The definition of the 2D discrete linear chirp transform is introduced and then it is used to construct the basis of the novel encryption algorithm. Finally, security analysis are performed to show the quality of the encryption process using different metrics.
\end{abstract} %


 \ifCLASSOPTIONpeerreview
 \begin{center} \bfseries EDICS Category: 3-BBND \end{center}
 \fi
\IEEEpeerreviewmaketitle

\section{Introduction}
\IEEEPARstart{W}{}ith the rapid development of digital technologies and Internet, information security including image encryption has become more and more important. The traditional encryption algorithms such as data encryption standard (DES), advanced encryption standard (AES), and public key encryption algorithm (RSA) are not suitable for image encryption because the digital image has intrinsic properties such as bulk data capacities, high redundancy, and strong correlation between the adjacent pixels \cite{Pareek2006, Yang2010}.  Therefore, many encryption methods relying on different approaches have been introduced in literature to fulfil the security requirements of digital images. Among these approaches, encryption algorithms  based on spatial domain \cite{Matthews1989,Zhu2011}, frequency domain \cite{Refregier1995,Luo2015}, and fractional domain \cite{Unnikrishnan2000} have been proposed for encryption algorithm design.

In spatial domain methods, the encryption algorithm works on the pixels of plain image directly, while the frequency domain schemes act on the coefficients of the transformed image which can be attained using transformation tools such as fast Fourier transform (FFT), discrete cosine transform (DCT), and discrete wavelet transform (DWT). On the other hand, the fractional domain approaches can give greater complexity to the system by giving an extra parameter of the transform order, which enlarges the key space resulting in a better and secure data
 protection as compared to the spatial and frequency domain methods. The most widely investigated fractional transform is the fractional Fourier transform, which has a well established continuous--time version and also several definitions in the discrete--time framework.

 In recent years, many different fractional Fourier transform encryption schemes have been proposed. In \cite{Zhu2000}, Zhu et al. proposed an optical image encryption method based on multi--fractional Fourier transforms (MFRFT).  Pei and Hsue presented an image encryption method based on multiple--parameter discrete fractional Fourier transform (MPDFRFT) \cite{Pei2006}.  In \cite{Liu2007}, Liu et al. introduced a random fractional Fourier transform (RFRFT) by using random phase modulations. The results in \cite{Ran2009} have shown that these image encryption schemes had deficiencies. Therefore, many encryption algorithms based on the FrFT  were suggested to enhance security \cite{Zhou2011,Lang2012,Elshamy2013,Sui2014,Li2015,Zhao2016}. %Although these encryption algorithms have good performance, the encrypted images are fully complex which makes the costs of transmission and storage high.

 In this paper, the definition of the 2D discrete linear chirp transform (2D-DLCT) is proposed and an image encryption scheme based on such transform is introduced. The 2D-DLCT is an extension to the 1D discrete linear chirp transform given in \cite{Osa2014}, which has an excellent property in a sense that the chirp rate parameter ideally can have  infinity support such that $-\infty< \beta <+\infty$  compared to the support of  the fractional order of the fractional Fourier transform $0\le \alpha \le \pi/2$. In the proposed encryption scheme, the plain image is scrambled using chaotic logistic map which has three secret keys. The scrambled image  is 2D-DLCT transformed using $\beta_x$ and $\beta_y$ chirp rates. These chirp rates serves as a secret keys as well. Then the transformed image is scrambled using different logistic map with another set of secret keys. Numerical  results show the proposed scheme can be infeasible to the brute--force attack, more secure, and  can resist  noise and occlusion attacks.%Finally, the transformed scrambled image is inversely converted back to the spatial domain where the encrypted image becomes real not complex such as the algorithms which based on the FrFT as illustrated before. Simulation results and a security analysis are used to demonstrate the security and feasibility of the proposed method.

The remainder of this paper is organized as follows. In section II, the definition of the 2D discrete Linear chirp transform is introduced. Section III presents the details of the proposed image encryption scheme. Section IV gives the numerical simulations and results to demonstrate the performance and verify the validity of the proposed scheme. The conclusion is drawn and stated in section V.

\section{2D Discrete Linear Chirp Transform (2D-DLCT)}
Let $x(n,m)$ be a two--dimensional discrete signal, where $n=0,1,\cdots,N-1$ and $m=0,1,\cdots,M-1$. The two--dimensional discrete linear chirp transform of the signal $x(n,m)$ with chirp rates $\beta_x$ for the x--axis and $\beta_y$ for the y--axis is defined as

\begin{eqnarray}
X(k,\ell)=\sum_{n=0}^{N-1}\sum_{m=0}^{M-1}x(n,m)~\mathcal{K}_{\beta_x,\beta_y}(n,m,k,\ell)
\label{eq1}
\end{eqnarray}
with the kernel
\begin{eqnarray}
\mathcal{K}_{\beta_x,\beta_y}(n,m,k,\ell)=\mathcal{K}_{\beta_x}(n,k)~\mathcal{K}_{\beta_y}(m,\ell)
\label{eq2}
\end{eqnarray}
where,
\begin{eqnarray}
\mathcal{K}_{\beta_x}(n,k)=\exp\left(-\frac{2\pi}{N}(k~n+\beta_x~n^2)\right)
\label{eq3}
\end{eqnarray}
and
\begin{eqnarray}
\mathcal{K}_{\beta_y}(m,\ell)=\exp\left(-\frac{2\pi}{M}(\ell~m+\beta_y~m^2)\right)
\label{eq4}
\end{eqnarray}
The chirp rates $\beta_x$ and $\beta_y$ are real numbers which can take any value from their support $-\infty<\beta_x,~\beta_y<\infty$. The 2D-DLCT can be expressed as a tensor product of two one--dimensional transforms. The inverse 2D-DLCT is obtained using the following mathematical expression
\begin{eqnarray}
x(n,m)=\sum_{k=0}^{N-1}\sum_{\ell=0}^{M-1}X(k,\ell)~\mathcal{K}_{\beta_x,\beta_y}^*(n,m,k,\ell)
\label{eq5}
\end{eqnarray}
where $0\le k \le N-1$, $0\le \ell \le M-1$, and ($*$) denotes the conjugate.

\section{The Proposed Encryption and Decryption Algorithm}
\begin{figure}[h]
\begin{minipage}[b]{0.5\linewidth}
\vspace{-0.5cm}
 % \centering
 \includegraphics[trim= 4cm 7.5cm 2cm 8.5cm, clip, width=10cm]{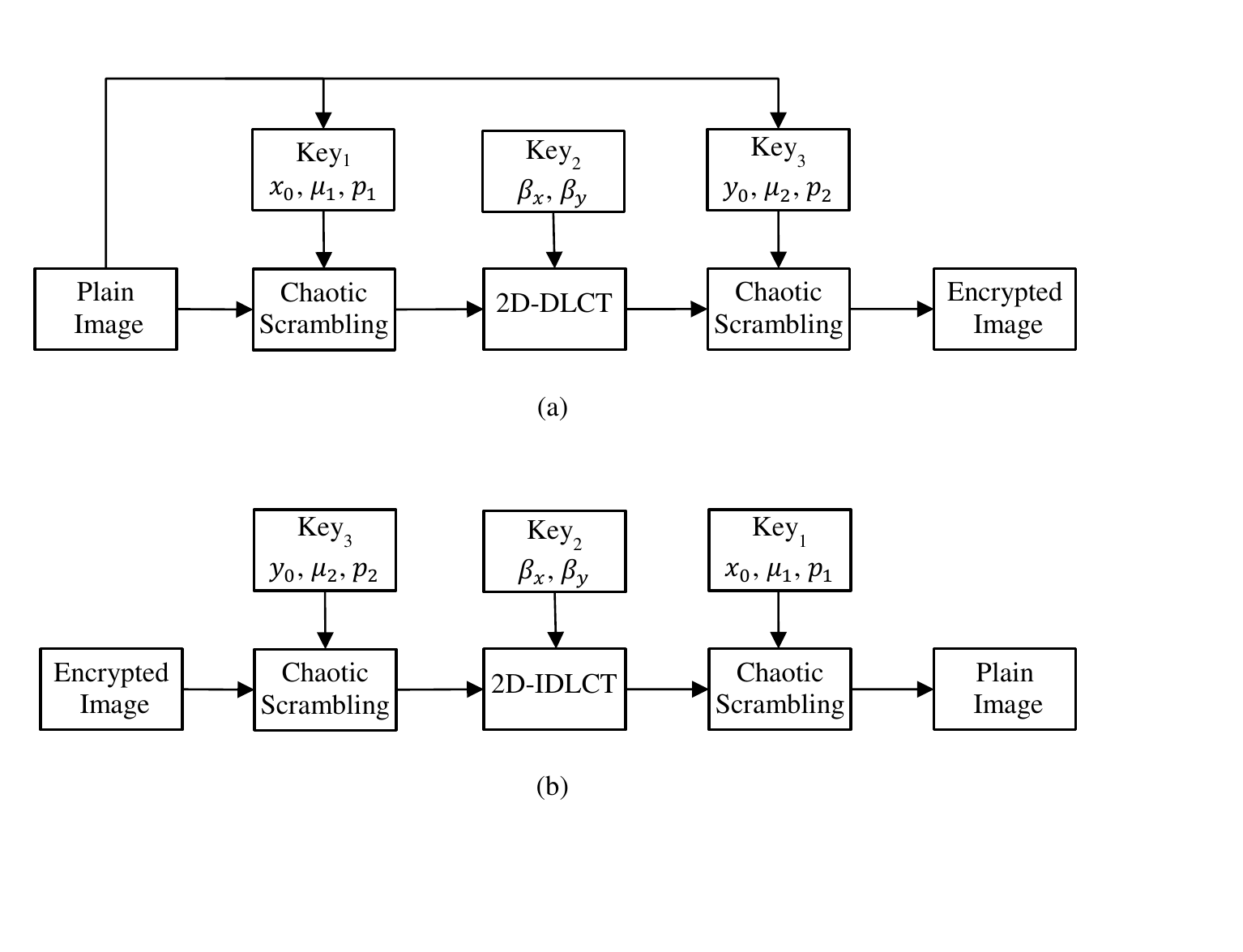}\\%fig8 fig23
  \end{minipage}
  \vspace{-1.5cm}
\caption{Schematic diagram of the proposed algorithm: (a) Encryption process, (b) Decryption process.}
\label{fig2}
\end{figure}

The proposed image encryption scheme is illustrated in Fig. \ref{fig2}(a). It employs the 2D-DLCT developed in Section II and  additional  strategies such as pixel rearrangement in the spatial and chirp rate domains using the well known chaotic logistic maps. The logistic map is a one--dimensional nonlinear chaos function and is defined as \cite{Matthews1989}
\begin{eqnarray}
x_{i+1}=\mu x_i(1-x_i)
\label{eq6}
\end{eqnarray}
where $\mu$ is the system parameter sometimes known as bifurcation parameter and $x_i\epsilon(0,1)$ is the sequence value. When $3.5699456\le \mu \le 4$, slight variations in the initial value yield dramatically different results over time. That is to say, logistic map will operate in chaotic state. With $x_0$ being the initial value, a non--periodic sequence $\{x_i; i=0,1,2,3,\cdots\}$ sensitive to the initial value is generated.

Without loss of generality, the size of the plain image I is an $N\times M$ and the encryption procedure is described as follows:
\begin{itemize}
  \item Setting the initial values of the chaotic system by means of the plain image $I$ to increase the relationship between the encryption scheme and the plain image.
  \item Given the initial parameter $x_0$ and $\mu_1$, generate a random sequence $x=\{x_1,x_2,\cdots,x_L\}$, where $L>N\times M$, and discard the first $p_1$ values. A new sequence $v=\{v_{1+p_1},v_{2+p_2},\cdots,v_{NM+p_1}\}$ is obtained. The parameters $x_0$, $\mu_1$, and $p_1$ serve as a first private key (key$_1$).
  \item Sort the sequence $v$ in ascending order or descending order to form a new sequence. Thus, the positions of the elements are varied and the positions are recorded as $IX$.
  \item Reshape the plain image $I_{N\times M}$ into a vector $S_{NM\times 1}$, and then use the scrambling index $IX$ to reorder the items of the vector $S$. The scrambled image in the spatial domain can be obtained by reconverting $S$ into an $N\times M$  matrix $E$.
  \item Choose two real numbers for $\beta_x$ and $\beta_y$, which serve as a second private key (key$_2$), and employ them to transform the image $E$ using 2D-DLCT.
  \item Scramble the attained matrix as explained in the previous steps using different set of initial values, that is  $y_0$, $\mu_2$, and $p_2$ which represent the third private key (key$_3$).
  \item Finally, the encrypted image is obtained by converting the scrambled vector into an $N\times M$ matrix $Y$.
\end{itemize}

The decryption algorithm is the inverse operation of the encryption as shown in Fig. \ref{fig2}(b).

\section{Simulation Results and Security Analysis}
In the simulations, a standard Lena test image of size $256\times 256$ \cite{imagedata} that has allocation of $8$ bits/pixel of gray--scale is used. The parameters of the chaotic logistic map and the chirp rates of the 2D-DLCT which employed in the simulation experiments are listed in Table \ref{Tab1}.
\begin{table}[h]
\caption{Simulation parameters.}
\label{Tab1}
\begin{center}
\begin{tabular} {| c | c | }
\hline
Parameter & Value  \\
\hline
$x_0$ & $0.31$  \\
\hline
$\mu_1$ & $3.8$ \\
\hline
$p_1$ & $\mod(\sum_{i,j}I_{ij},9999)$ \\
\hline
$y_0$ & $0.25$ \\
\hline
$\mu_2$ & $3.7$ \\
\hline
$p_2$ & $\mod(\sum_{i,j}I_{ij},9990)$ \\
\hline
$\beta_x$ & $1.5$  \\
\hline
$\beta_y$ & $-3.5$ \\
\hline
\end{tabular}
\end{center}
\end{table}

\begin{figure}[h]
\begin{minipage}[b]{0.5\linewidth}
%\vspace{-0.5cm}
  \centering
 \includegraphics[trim= 7cm 10cm -4cm 10cm, clip, width=11cm]{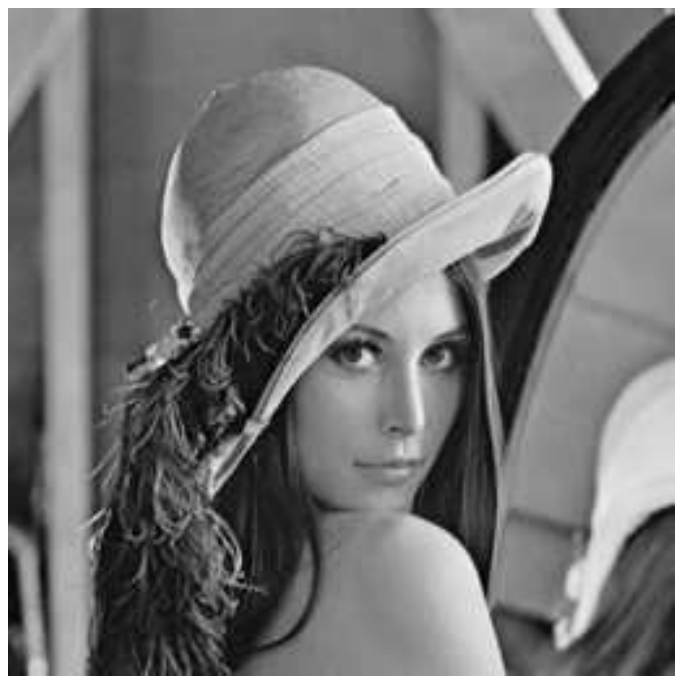}\\
 \vspace{-0.5cm}
  \centerline{(a)}
  \end{minipage}%
\begin{minipage}[b]{0.5\linewidth}
%\vspace{-0.5cm}
  \centering
\includegraphics[trim= 7cm 10cm -4cm 10cm, clip, width=11cm]{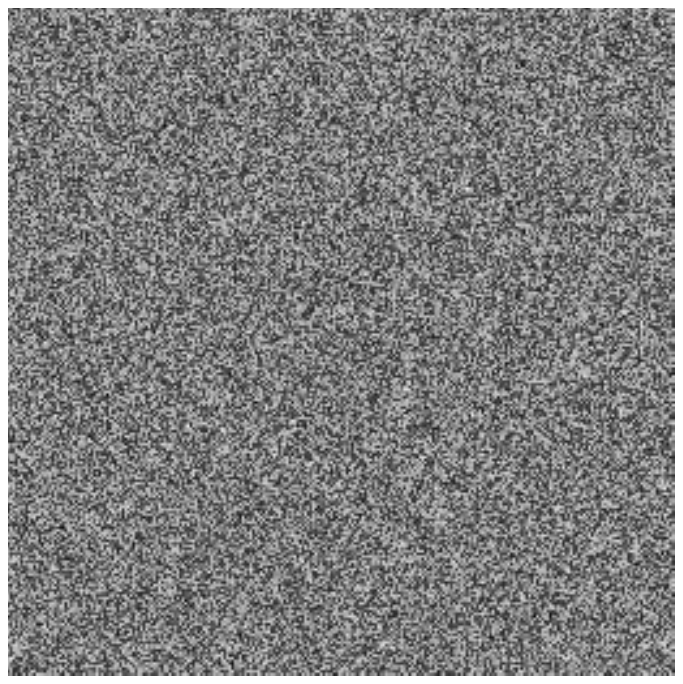}\\
 \vspace{-0.5cm}
  \centerline{(b)}
  \end{minipage}
  \begin{minipage}[b]{0.5\linewidth}
  %\vspace{0.cm}
  \centering
  \includegraphics[trim= 7cm 10cm -4cm 10cm, clip, width=11cm]{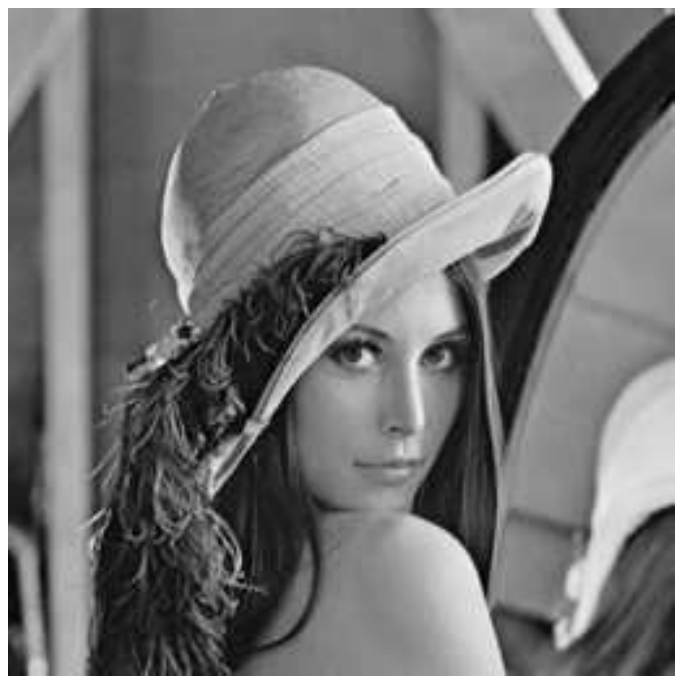}\\
  \vspace{-0.5cm}
  \centerline{(c)}
 \end{minipage}
\caption{Encryption and decryption results. (a) Plain image. (b) Encrypted image. (c) Decrypted image.}
\label{fig3}
\end{figure}

Figures \ref{fig3}(a), (b), and (c) show the plain image, encrypted image, and decrypted image, respectively. In the following subsections, the security analysis of the proposed encryption scheme is performed to check its resistance to various attacks.

\subsection{Key Space Analysis}
Key space size is the total number of different keys that can be used in an encryption algorithm. In a cryptographic system, the key space should be sufficiently large to make brute--force attack infeasible.  The proposed encryption algorithm  have the following secret keys: key$_1=\{x_0,\mu_1,p_1\}$, key$_2=\{\beta_x, \beta_y\}$ , and  key$_3=\{y_0,\mu_2,p_2\}$  and their corresponding spaces are $s_1$, $s_2$, $s_3$, $s_4$, $s_5$, $s_6$, $s_7$, and $s_8$ , respectively. These keys are independent from each other. Thus the total key space of the encryption scheme can be computed as
\begin{eqnarray}
s>\prod_{i}s_i
\label{eq7}
\end{eqnarray}

If we assume the computation precision of the computer is $10^{-14}$, then the key space is about $10^{74}\approx 2^{245}$. Such a large key space can ensure a high security against brute--force attacks \cite{Alvarez06}.

\subsection{Key Sensitivity Analysis}
 A good encryption scheme should be sensitive to each secret key. In other words, a small change on the key must be able to cause a great change on the encrypted image. In order to evaluate the key sensitivity of the proposed algorithm, the mean square error (MSE) between the plain image and decrypted image is calculated as follows
\begin{eqnarray}
\mbox{MSE}=\frac{1}{N\times M} \sum_{i=1}^N\sum_{j=1}^M |I(i,j)-D(i,j)|^2
\label{eq8}
\end{eqnarray}
where $I(i,j)$ is the plain image and $D(i,j)$ denotes the corresponding decrypted image. To determine the sensitivity of the key parameters, the decryption procedure is processed by varying one parameter while the others held constant.

\begin{figure}[h]
\begin{minipage}[b]{0.5\linewidth}
%\vspace{-0.5cm}
  \centering
 \includegraphics[trim= 3.6cm 7cm 3.6cm 9cm, clip, width=5cm]{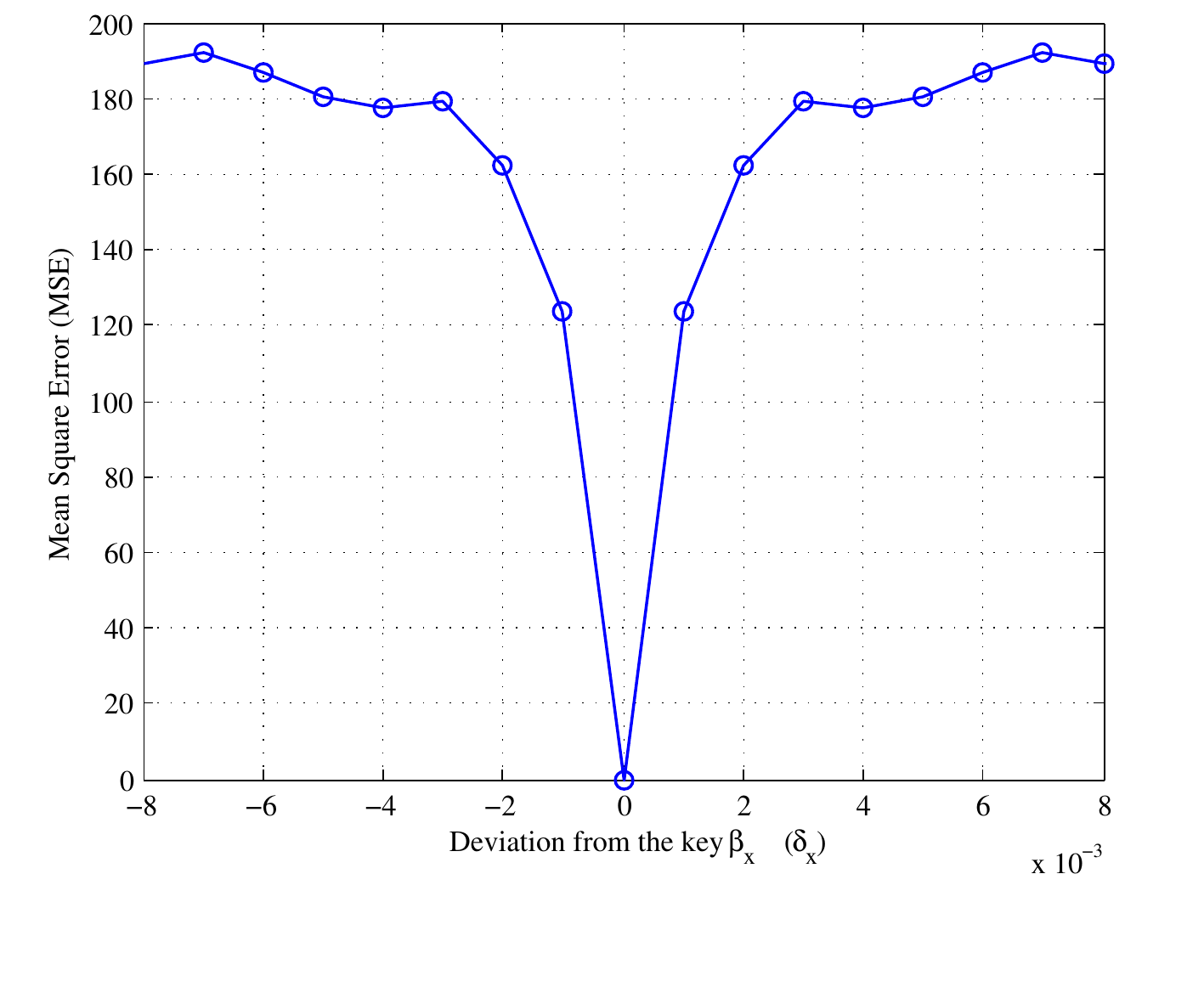}\\
 \vspace{-0.5cm}
  \centerline{(a)}
  \end{minipage}%
\begin{minipage}[b]{0.5\linewidth}
%\vspace{-0.5cm}
  \centering
\includegraphics[trim= 3.6cm 7cm 3.6cm 9cm, clip, width=5cm]{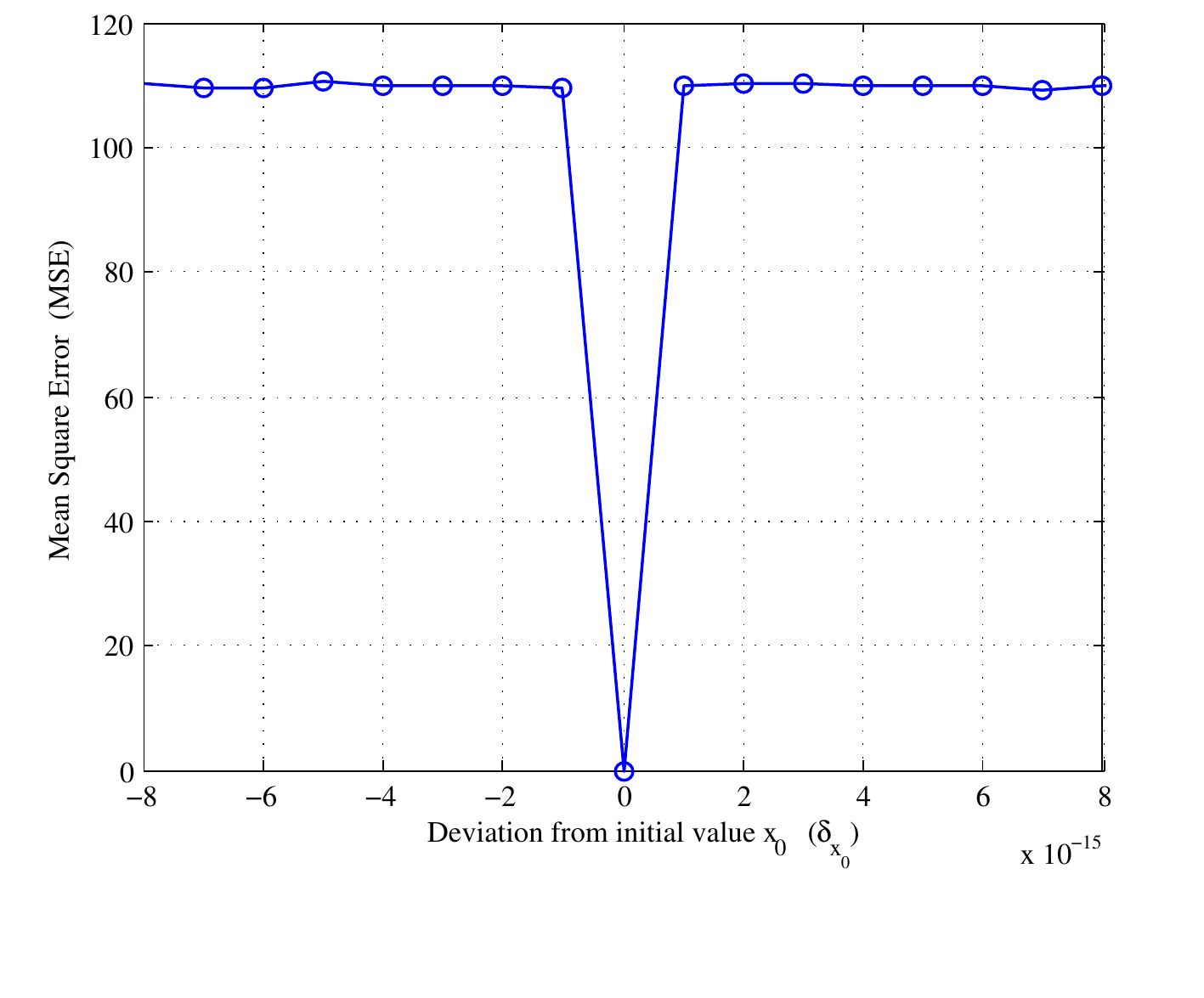}\\
 \vspace{-0.5cm}
  \centerline{(b)}
  \end{minipage}
  \begin{minipage}[b]{0.5\linewidth}
  %\vspace{0.cm}
  \centering
  \includegraphics[trim= 3.6cm 7cm 3.6cm 9cm, clip, width=5cm]{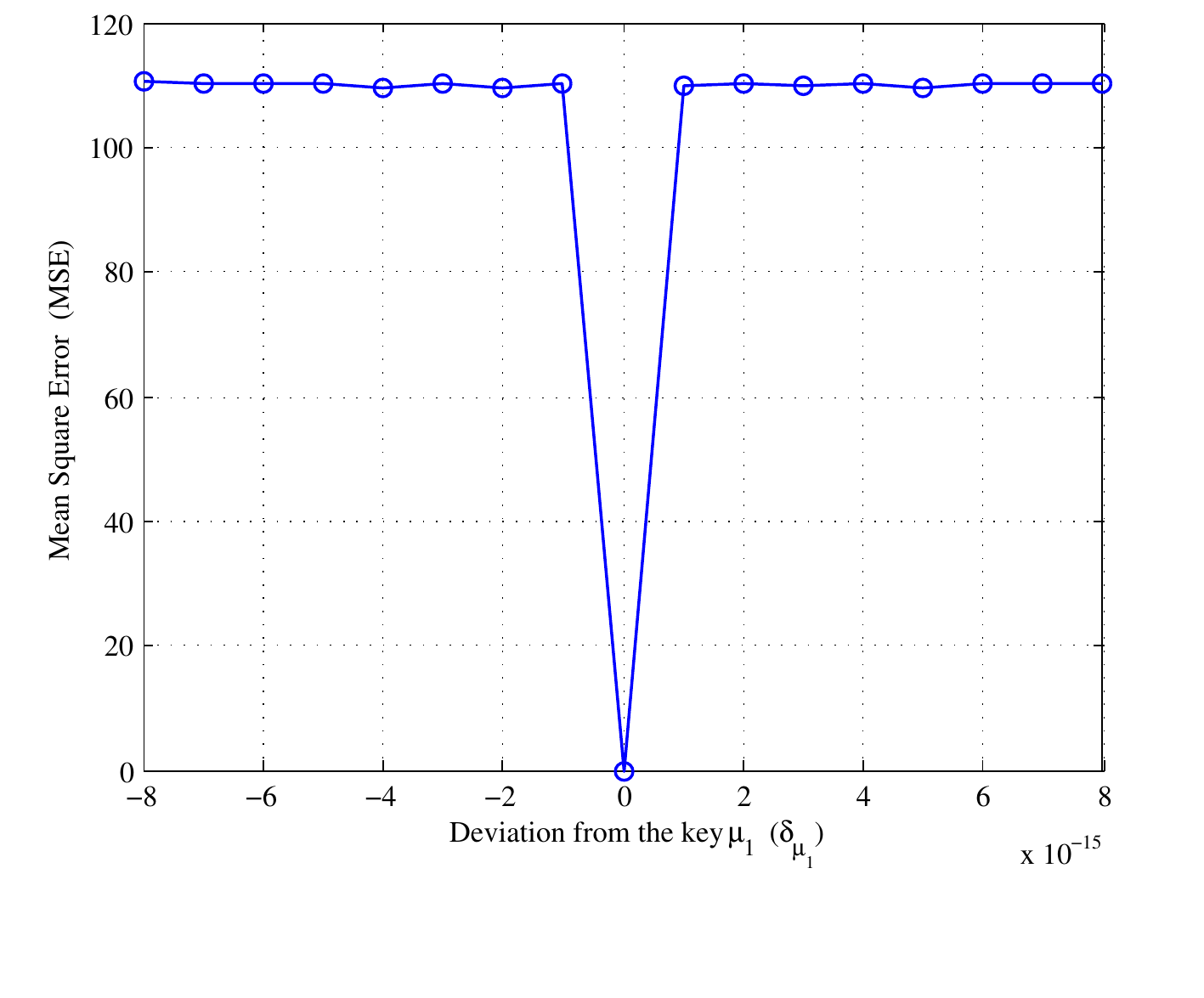}\\
  \vspace{-0.5cm}
  \centerline{(c)}
 \end{minipage}%
\begin{minipage}[b]{0.5\linewidth}
%\vspace{-0.5cm}
  \centering
\includegraphics[trim= 3.6cm 7cm 3.6cm 9cm, clip, width=5cm]{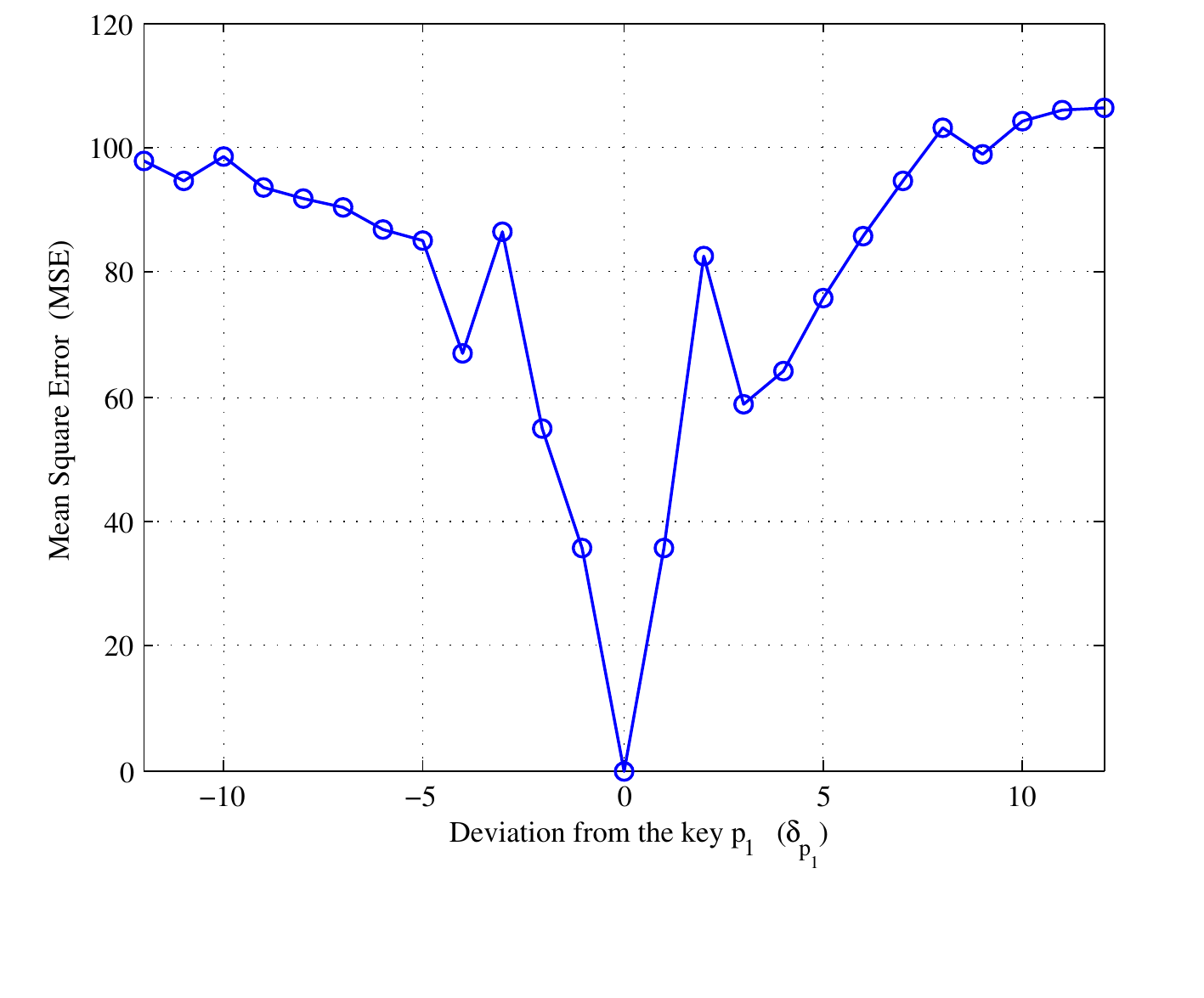}\\
 \vspace{-0.5cm}
  \centerline{(d)}
  \end{minipage}
\caption{MSE for deviation of the correct keys. (a) Chirp rate $\beta_x$. (b) Initial value $x_0$. (c) Logistic map function parameter $\mu_1$. (d) Truncated position $p_1$.}
\label{fig4}
\end{figure}

Figures \ref{fig4}(a), (b), (c), and (d) show the MSE versus deviation of the key encryption parameters $\beta_x$, $x_0$, $\mu_1$, and $p_1$, respectively. Since the MSE increases sharply when the the key parameters depart from its correct value and it is equal to zero when the image is decrypted with correct decryption parameters, a small change to the keys can lead to different decrypted image having no connection with the original image.

It should be noted that we omit the sensitivity analysis of the remaining key parameters $y_0$, $\mu_2$, and $p_2$ because they give similar results to those shown with the key parameters $x_0$, $\mu_1$, and $p_1$.

\begin{figure}[h]
\begin{minipage}[b]{0.5\linewidth}
%\vspace{-0.5cm}
  \centering
 \includegraphics[trim= 3.6cm 7cm 3.6cm 9cm, clip, width=5cm]{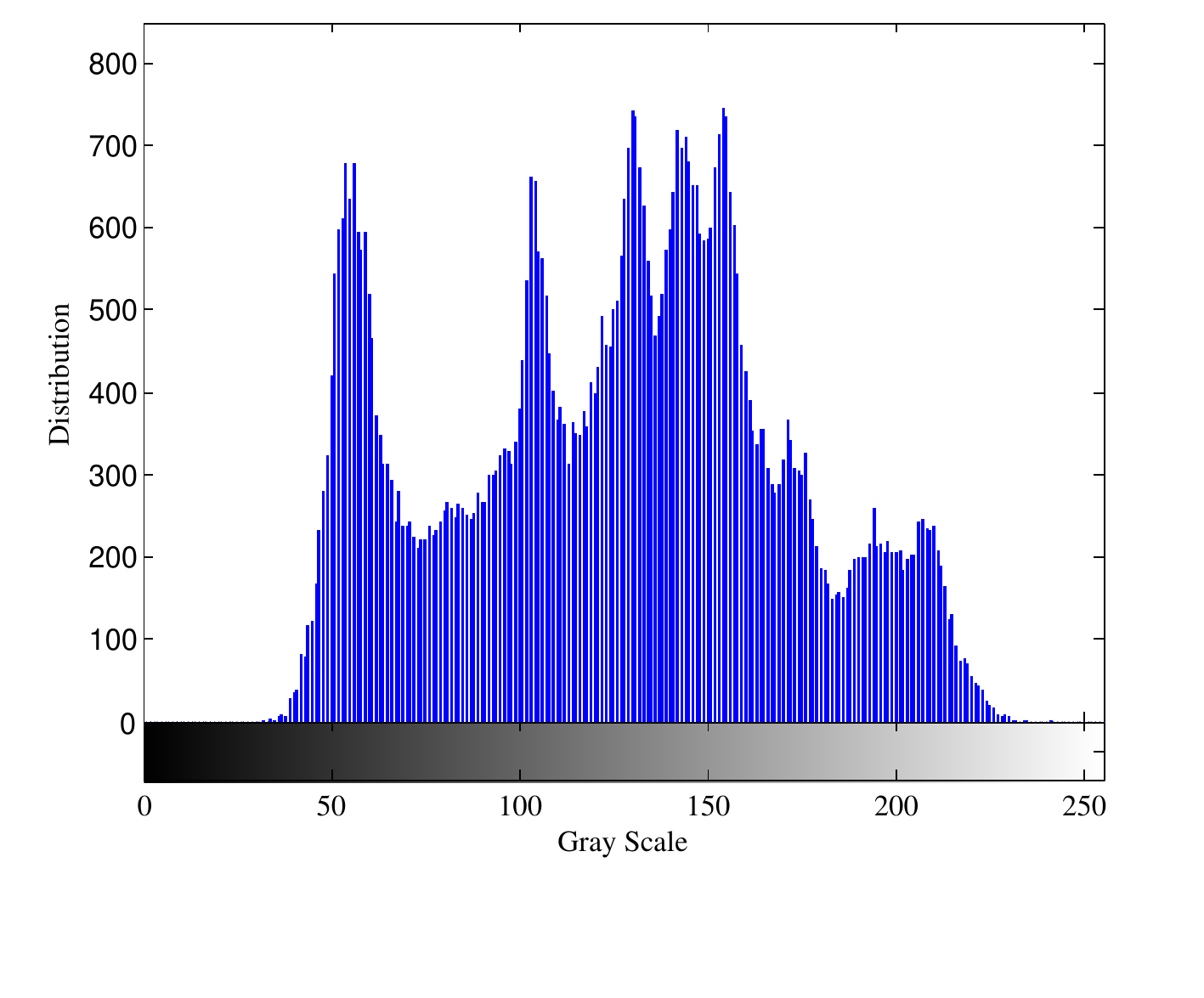}\\
 \vspace{-0.5cm}
  \centerline{(a)}
  \end{minipage}%
\begin{minipage}[b]{0.5\linewidth}
%\vspace{-0.5cm}
  \centering
\includegraphics[trim= 3.6cm 7cm 3.6cm 9cm, clip, width=5cm]{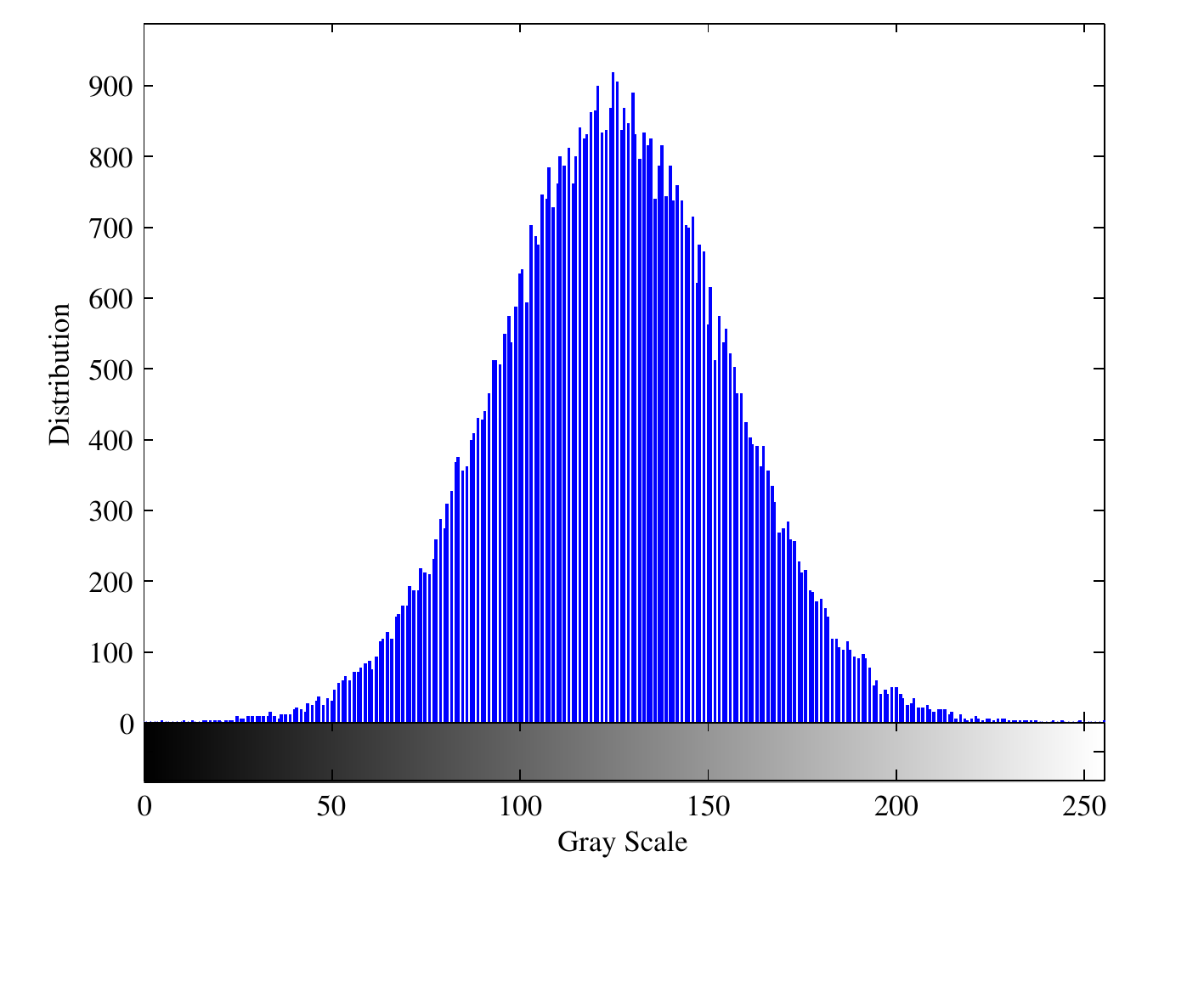}\\
 \vspace{-0.5cm}
  \centerline{(b)}
  \end{minipage}
  \begin{minipage}[b]{0.5\linewidth}
  %\vspace{0.cm}
  \centering
  \includegraphics[trim= 3.6cm 7cm 3.6cm 9cm, clip, width=5cm]{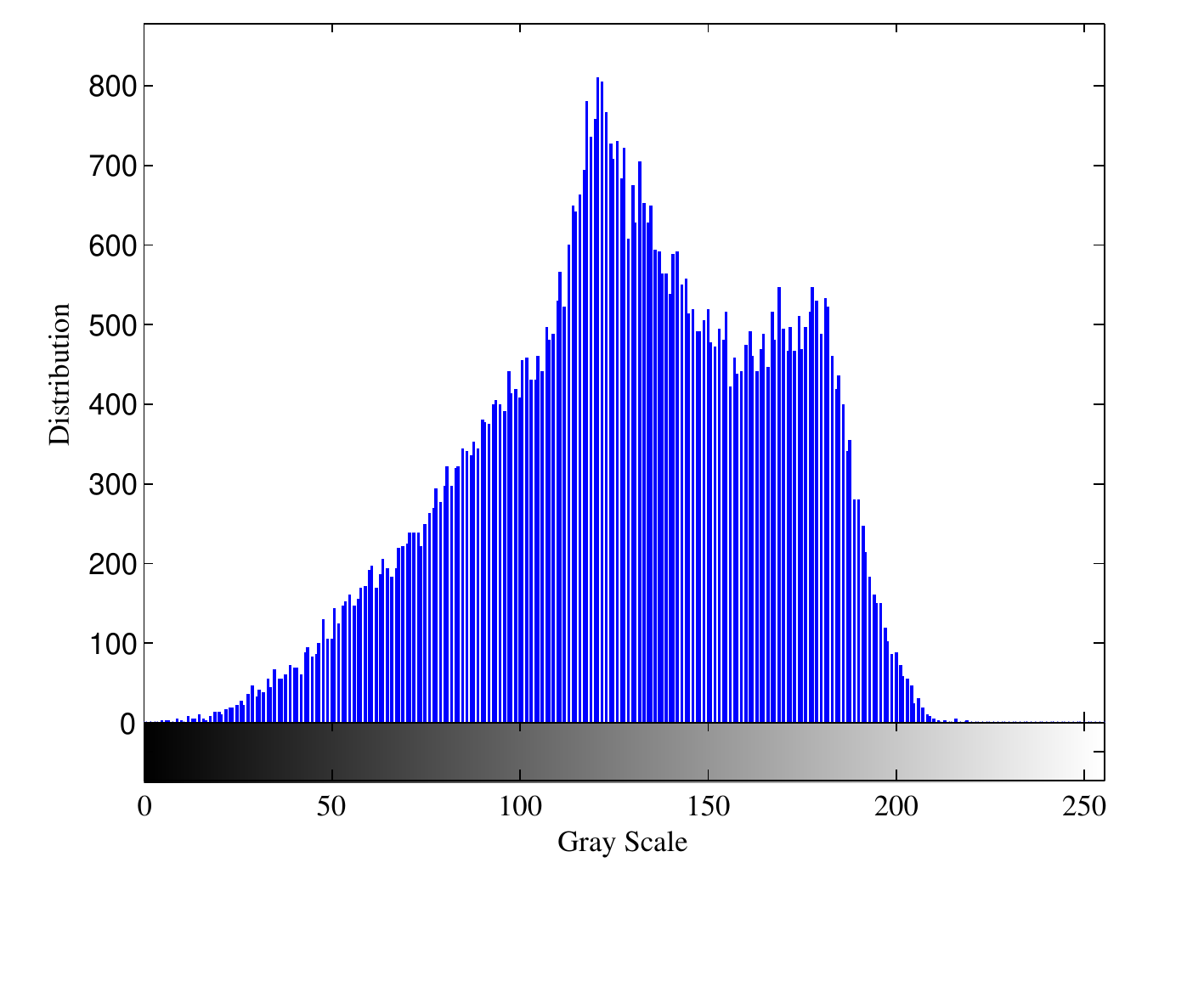}\\
  \vspace{-0.5cm}
  \centerline{(c)}
 \end{minipage}%
\begin{minipage}[b]{0.5\linewidth}
%\vspace{-0.5cm}
  \centering
\includegraphics[trim= 3.6cm 7cm 3.6cm 9cm, clip, width=5cm]{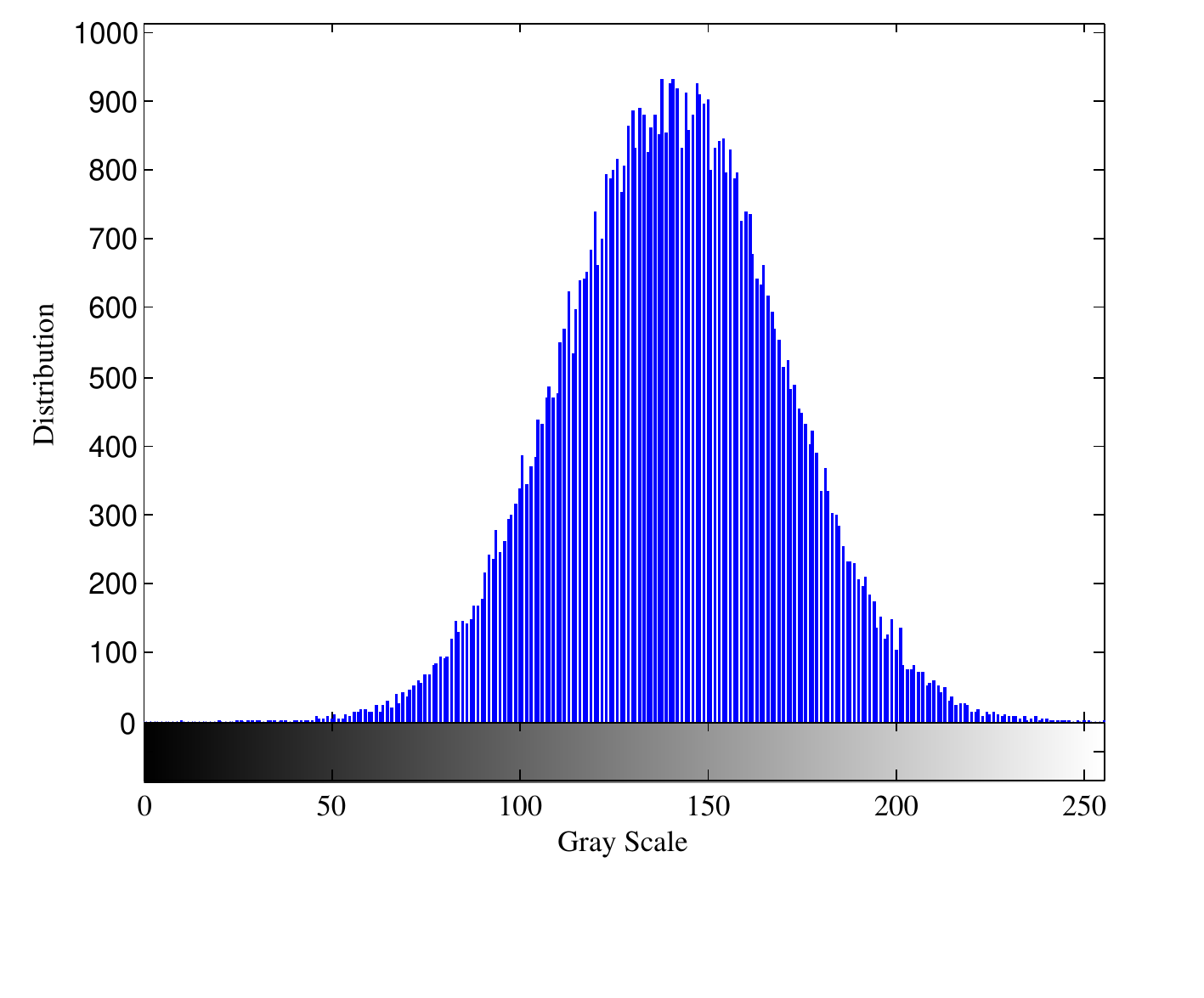}\\
 \vspace{-0.5cm}
  \centerline{(d)}
  \end{minipage}
\caption{Examples of histograms. (a) Lena image. (b) Encrypted image of Lena. (c) Baboon image. (d) Encrypted image of Baboon.}
\label{fig6}
\end{figure}

\subsection{Histogram Analysis}
To resist statistical attacks, the encrypted images should have  histograms that are consistent in distribution and are different from the histograms of their plain images. Figure \ref{fig6}(a) and (c) show the histograms of Lena and Baboon plain images respectively, while Fig. \ref{fig6}(b) and (d) present the histograms of their encrypted images. It is clear that the histograms of the plain images are different from each other and their corresponding encrypted images have similar statistical properties. Moreover, the values of the encrypted image are subject to normal distribution. Hence, the encrypted image histogram data does not provide any useful information for the attackers to perform any statistical analysis attack on the encrypted image.

\subsection{Correlation Analysis}
\begin{figure}[h]
\begin{minipage}[b]{0.5\linewidth}
%\vspace{-0.5cm}
  \centering
 \includegraphics[trim= 3.6cm 7cm 3.6cm 9cm, clip, width=5cm]{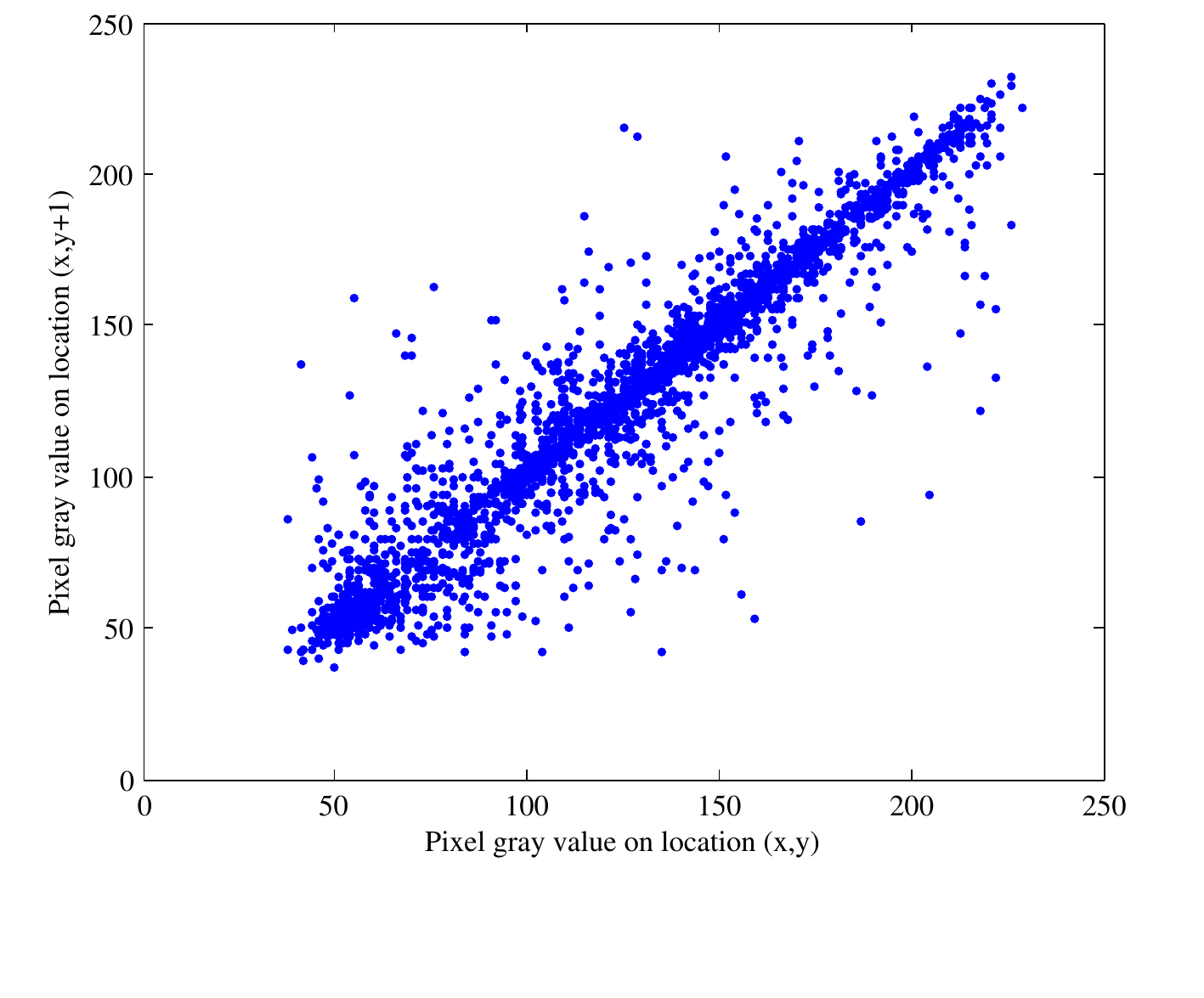}\\
 \vspace{-0.5cm}
  \centerline{(a)}
  \end{minipage}%
\begin{minipage}[b]{0.5\linewidth}
%\vspace{-0.5cm}
  \centering
\includegraphics[trim= 3.6cm 7cm 3.6cm 9cm, clip, width=5cm]{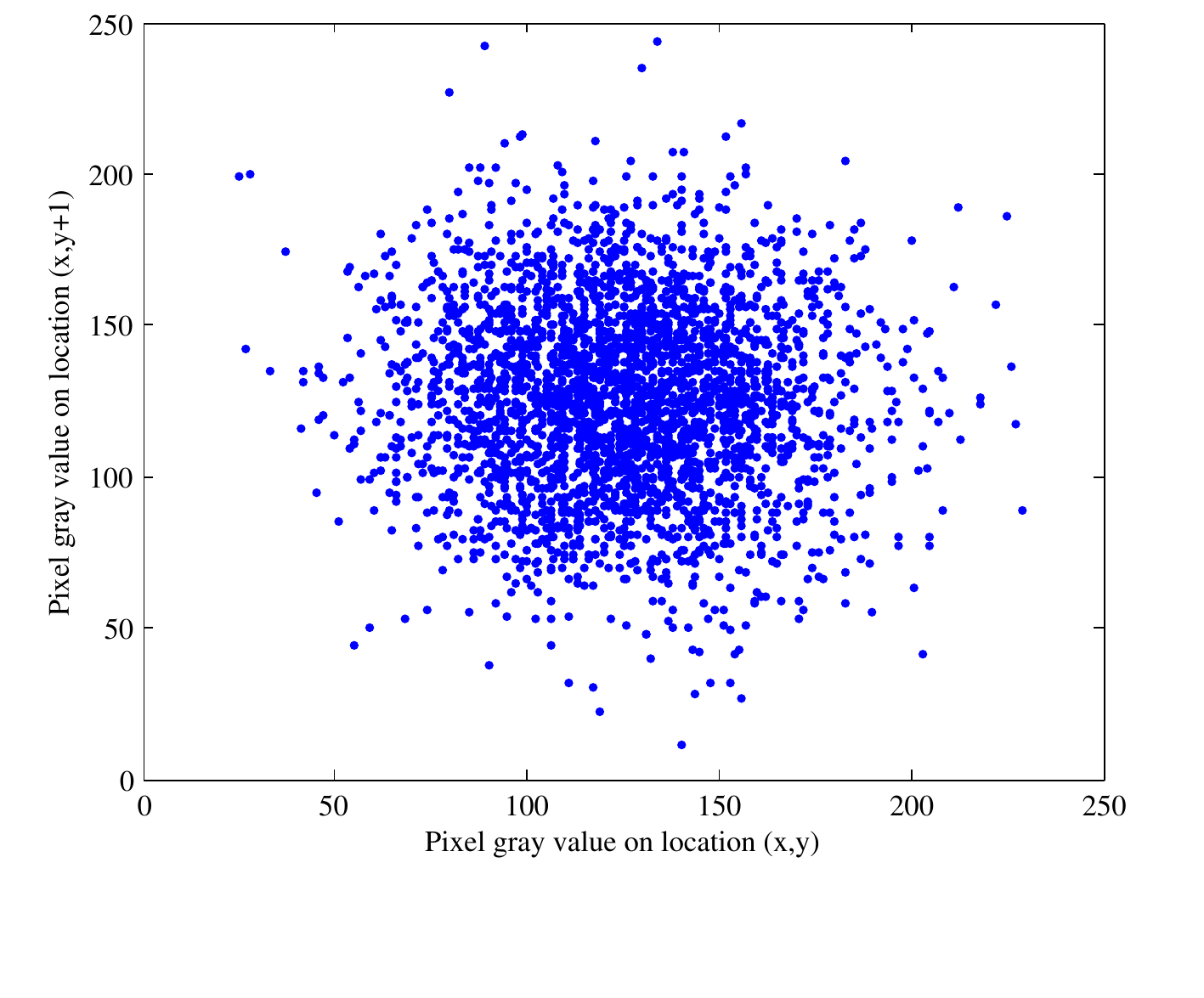}\\
 \vspace{-0.5cm}
  \centerline{(b)}
  \end{minipage}
  \begin{minipage}[b]{0.5\linewidth}
  %\vspace{0.cm}
  \centering
  \includegraphics[trim= 3.6cm 7cm 3.6cm 9cm, clip, width=5cm]{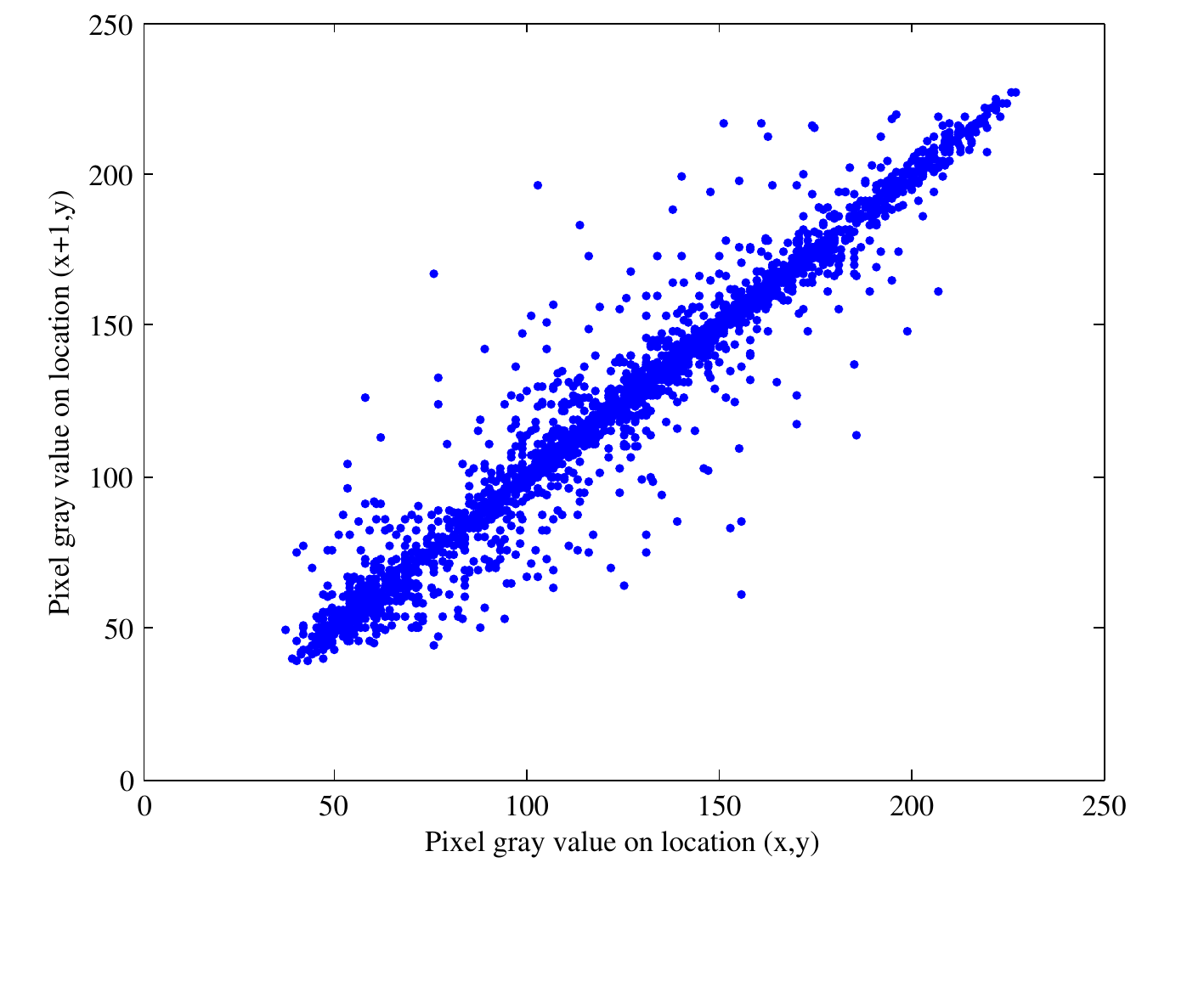}\\
  \vspace{-0.5cm}
  \centerline{(c)}
 \end{minipage}%
\begin{minipage}[b]{0.5\linewidth}
%\vspace{-0.5cm}
  \centering
\includegraphics[trim= 3.6cm 7cm 3.6cm 9cm, clip, width=5cm]{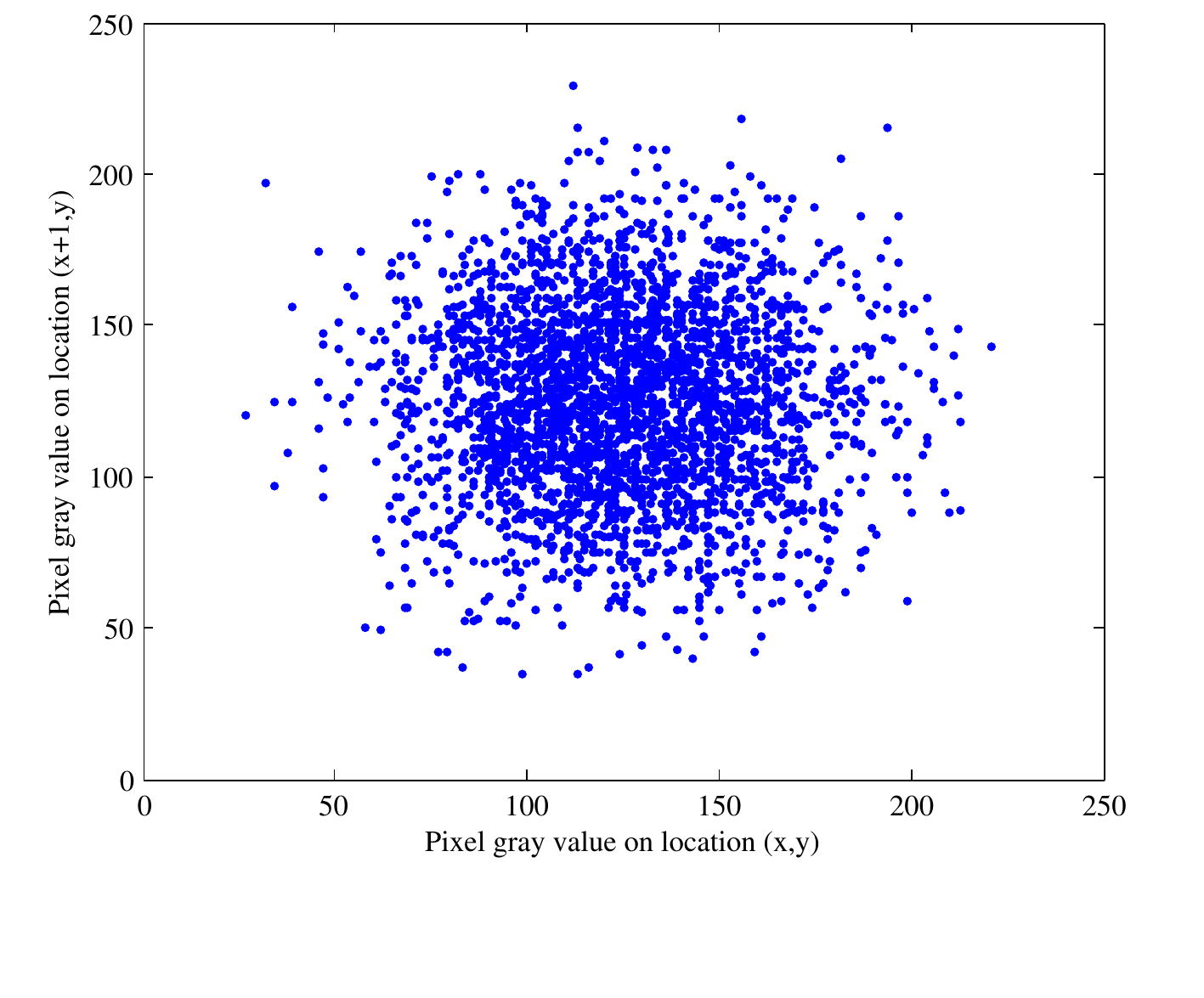}\\
 \vspace{-0.5cm}
  \centerline{(d)}
  \end{minipage}
  \begin{minipage}[b]{0.5\linewidth}
  %\vspace{0.cm}
  \centering
  \includegraphics[trim= 3.6cm 7cm 3.6cm 9cm, clip, width=5cm]{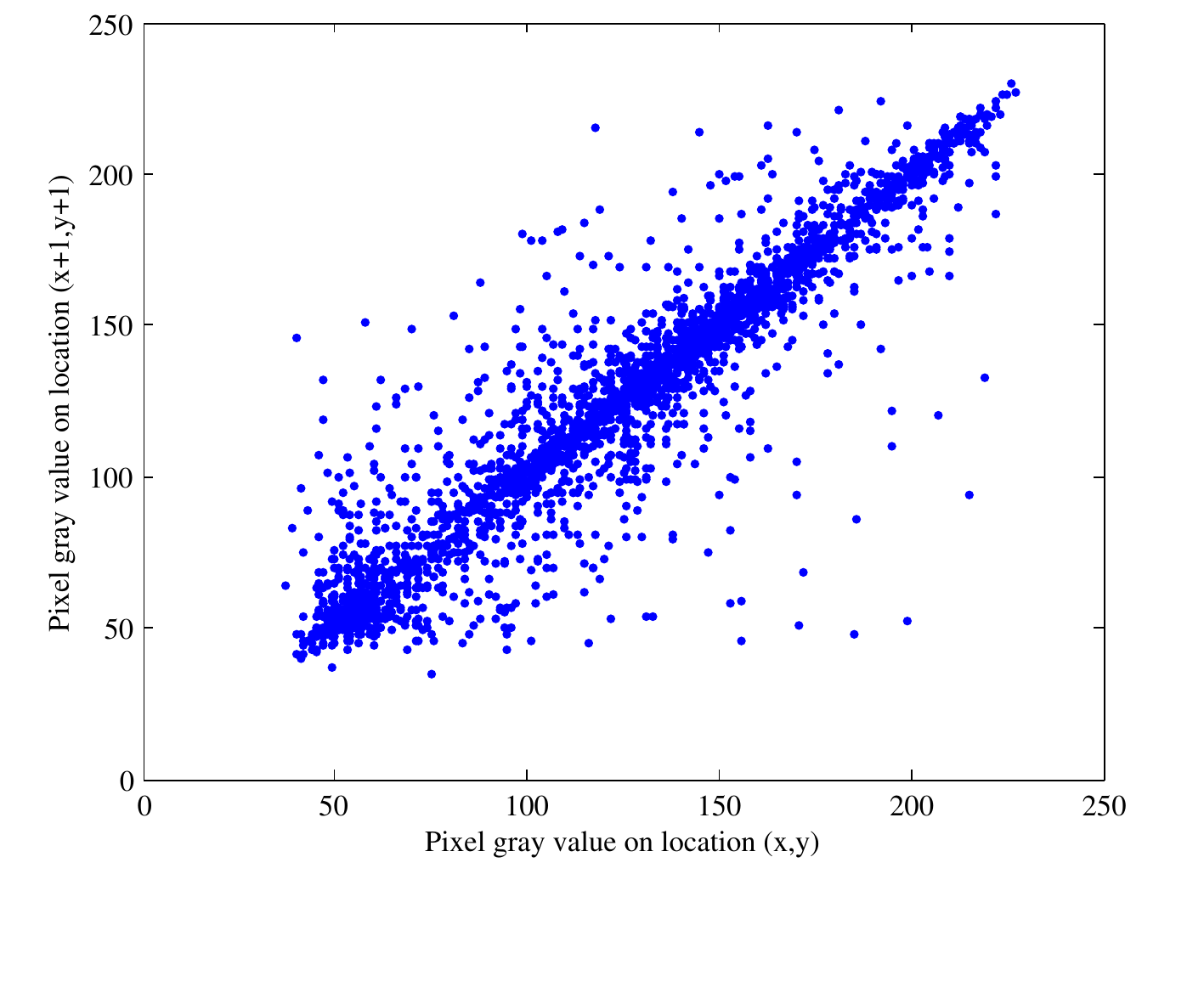}\\
  \vspace{-0.5cm}
  \centerline{(e)}
 \end{minipage}%
\begin{minipage}[b]{0.5\linewidth}
%\vspace{-0.5cm}
  \centering
\includegraphics[trim= 3.6cm 7cm 3.6cm 9cm, clip, width=5cm]{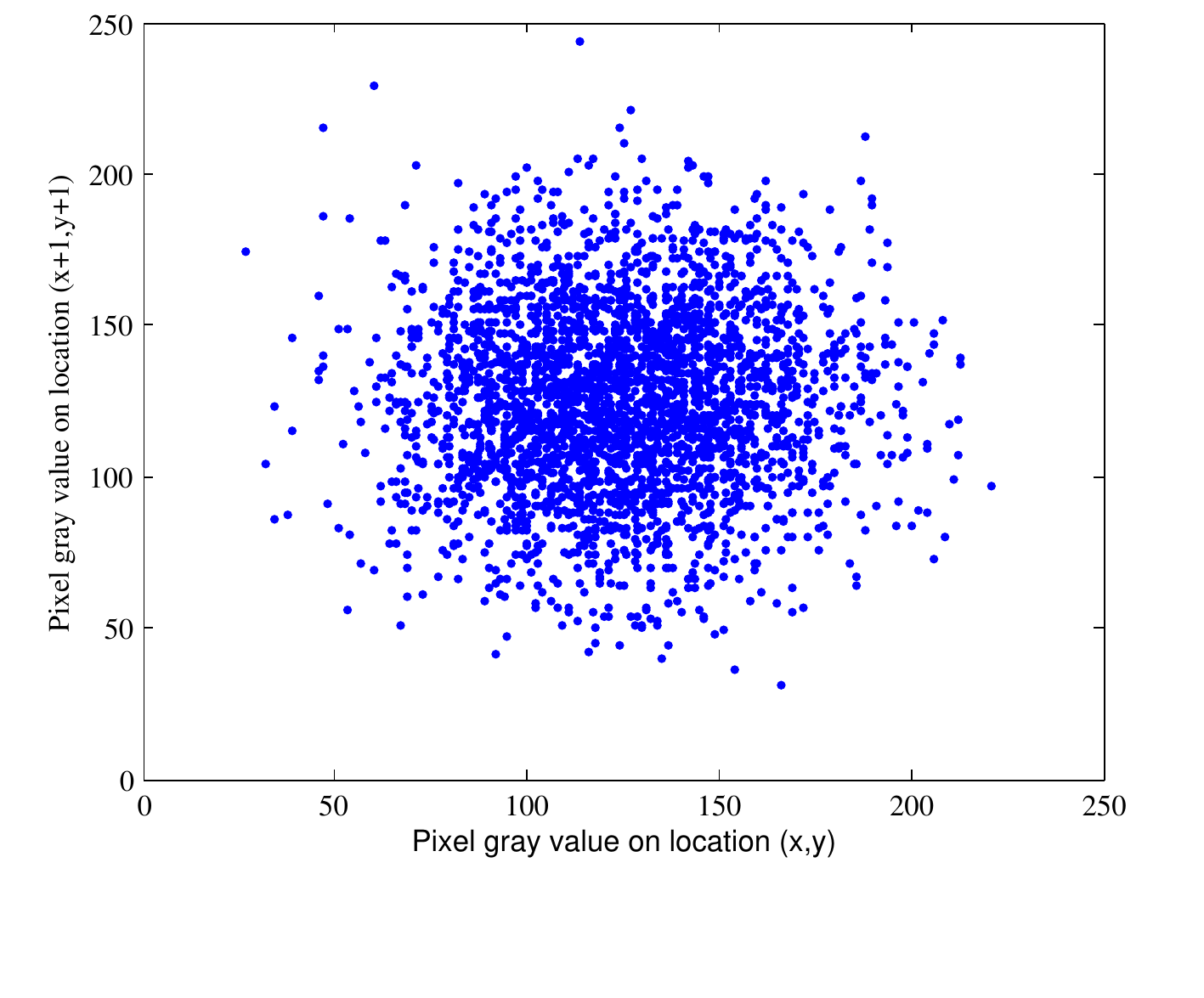}\\
 \vspace{-0.5cm}
  \centerline{(f)}
  \end{minipage}
\caption{Correlation coefficients of plain and encrypted Lena images, from top to bottom: in horizontal, vertical, and diagonal direction, respectively.}
\label{fig7}
\end{figure}

To evaluate the correlation of adjacent pixels, $6000$ pairs of adjacent pixels are randomly selected from both the plain and encrypted images. A good encrypted image must have low correlation for the three directions-– horizontal, vertical and diagonal. The distribution of two adjacent pixels to the plain and encrypted images for the three directions is shown in Fig. \ref{fig7}(a)–(f). Thus, correlation plots of plain images exhibit clear patterns while those of their encrypted counterparts show no clear pattern, the points being rather randomly distributed.

In order to test the correlations of adjacent pixels for the plain and encrypted images, the correlation coefficients $r_{xy}$ of each pair are calculated using the following equations
\begin{eqnarray}
r_{xy}=\frac{\sum_{i=1}^{L}(x_i-\bar{x})(y_i-\bar{y})}{\sqrt{\left[\sum_{i=1}^{L}(x_i-\bar{x})^2\right]\left[\sum_{i=1}^{L}(y_i-\bar{y})^2\right]}}
\label{eq9}
\end{eqnarray}
where $\bar{x}=1/L\sum_{i=1}^{L}x_i$, $\bar{y}=1/L\sum_{i=1}^{L}y_i$, $x_i$ is the value of the i-th selected pixel, $y_i$ is the value of the correspondent adjacent pixel, and $L$ is the total number of pixels selected from the image. The correlation coefficients of the plain image and encrypted image are summarized in Table \ref{Tab2}.

\begin{table}[h]
\caption{Correlation coefficients of two adjacent pixels in the plain and encrypted images.}
\label{Tab2}
\begin{center}
\begin{tabular} {| c | c | c | c | }
\hline
Image and scheme  & Horizontal & Vertical & Diagonal \\
\hline
   Plain Lena & $0.9481$ & $0.9737$ & $0.9244$\\
   \hline
   Encrypted Lena & $-0.0014$ & $0.0025$ & $0.0051$\\
   \hline
   Plain Baboon & $0.8701$ & $0.8411$ & $0.7889$\\
   \hline
   Encrypted Baboon & $0.0077$ & $-0.002$ & $-0.0039$\\
   \hline
   Plain Cameraman & $0.9556$ & $0.9738$ & $0.934$\\
   \hline
   Encrypted Cameraman & $-5.84\times 10^{-4}$ & $-0.0077$ & $0.0015$\\
   \hline
   Plain Pirate & $0.9434$ & $0.9564$ & $0.9134$\\
   \hline
   Encrypted Pirate & $-0.002$ & $5.25\times 10^{-4}$ & $-0.0048$\\
\hline
\end{tabular}
\end{center}
\end{table}

The importance of this work is also emphasised through the values of the correlation coefficient in Table \ref{Tab3}. It is clear that the correlation coefficient of the proposed
algorithm is smaller than that of other methods proposed in Refs. \cite{Lima2014,Sui2014,Nini2015,Annaby2016,Chai2017,Xu2017}.
\begin{table}[h]
\caption{Correlation coefficient of the proposed algorithm compared to others using plain Lena image.}
\label{Tab3}
\begin{center}
\begin{tabular} {| c | c | c | c | }
\hline
Algorithm  & Horizontal & Vertical & Diagonal \\
\hline

   Proposed algorithm & $-0.0014$ & $0.0025$ & $0.0051$\\
   \hline
   Ref. \cite{Lima2014} & $0.002$ & $0.0062$ & $0.0066$\\
   \hline
   Ref. \cite{Sui2014} & $-0.0069$ & $-0.0188$ & $-0.0482$\\
   \hline
   Ref. \cite{Nini2015} & $-0.0274$ & $0.0046$ & $-0.0038$\\
   \hline
   Ref. \cite{Annaby2016} & $0.319$ & $0.3314$ & $0.0149$\\
   \hline
   Ref. \cite{Chai2017} & $-0.0016 $ & $-0.0033$ & $0.0130 $\\
   \hline
   Ref. \cite{Xu2017} & $-0.0226$ & $0.0041$ & $0.0368$\\
\hline
\end{tabular}
\end{center}
\end{table}

\subsection{Occlusion Attack Analysis}
\begin{figure}[h]
\begin{minipage}[b]{0.5\linewidth}
%\vspace{-0.5cm}
  \centering
 \includegraphics[trim= 7cm 10cm -4cm 10cm, clip, width=11cm]{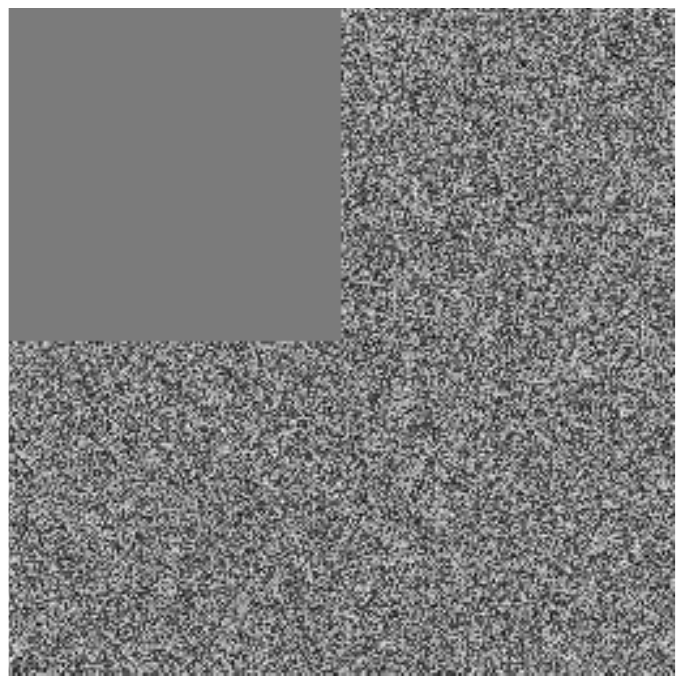}\\
 \vspace{-0.5cm}
  \centerline{(a)}
  \end{minipage}%
\begin{minipage}[b]{0.5\linewidth}
%\vspace{-0.5cm}
  \centering
\includegraphics[trim= 7cm 10cm -4cm 10cm, clip, width=11cm]{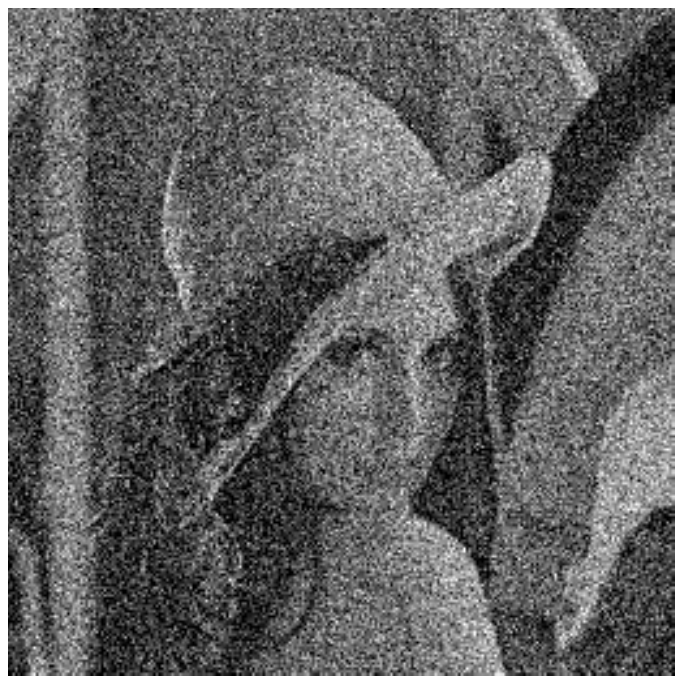}\\
 \vspace{-0.5cm}
  \centerline{(b)}
  \end{minipage}
  \begin{minipage}[b]{0.5\linewidth}
  %\vspace{0.cm}
  \centering
  \includegraphics[trim= 7cm 10cm -4cm 10cm, clip, width=11cm]{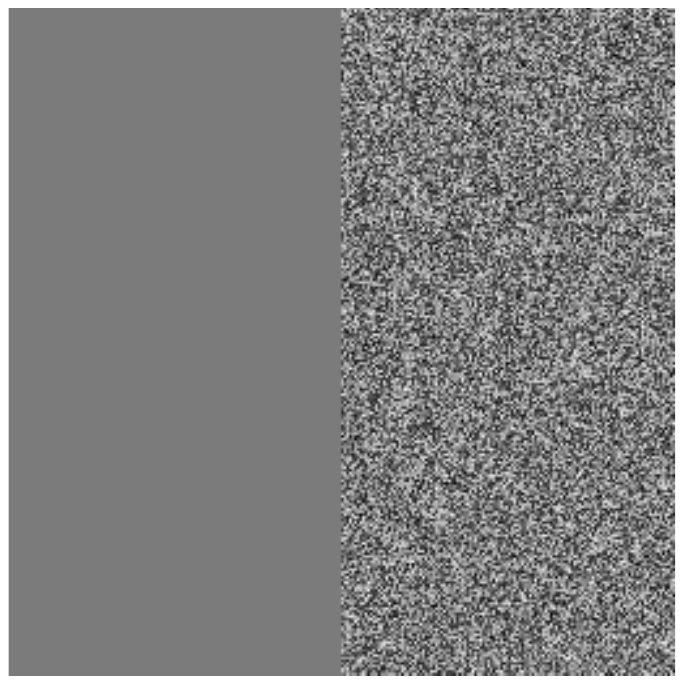}\\
  \vspace{-0.5cm}
  \centerline{(c)}
 \end{minipage}%
\begin{minipage}[b]{0.5\linewidth}
%\vspace{-0.5cm}
  \centering
\includegraphics[trim= 7cm 10cm -4cm 10cm, clip, width=11cm]{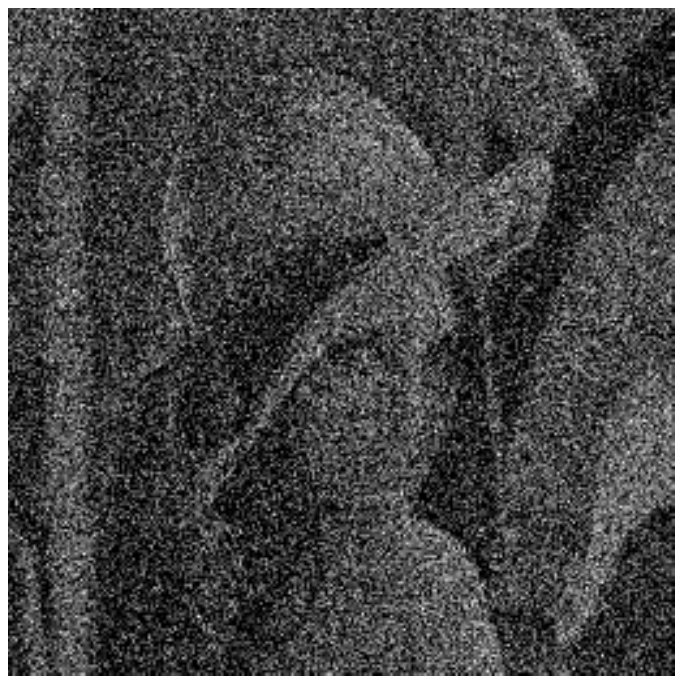}\\
 \vspace{-0.5cm}
  \centerline{(d)}
  \end{minipage}
  \begin{minipage}[b]{0.5\linewidth}
  %\vspace{0.cm}
  \centering
  \includegraphics[trim= 7cm 10cm -4cm 10cm, clip, width=11cm]{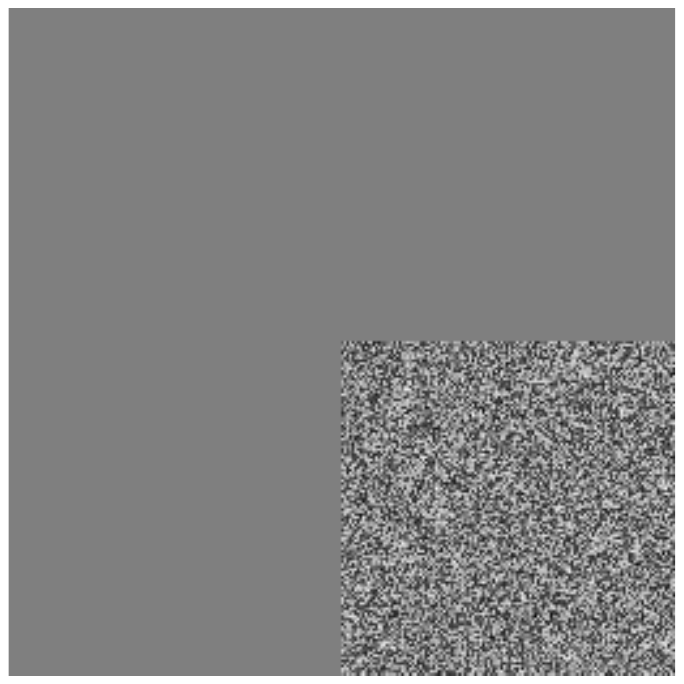}\\
  \vspace{-0.5cm}
  \centerline{(e)}
 \end{minipage}%
\begin{minipage}[b]{0.5\linewidth}
%\vspace{-0.5cm}
  \centering
\includegraphics[trim= 7cm 10cm -4cm 10cm, clip, width=11cm]{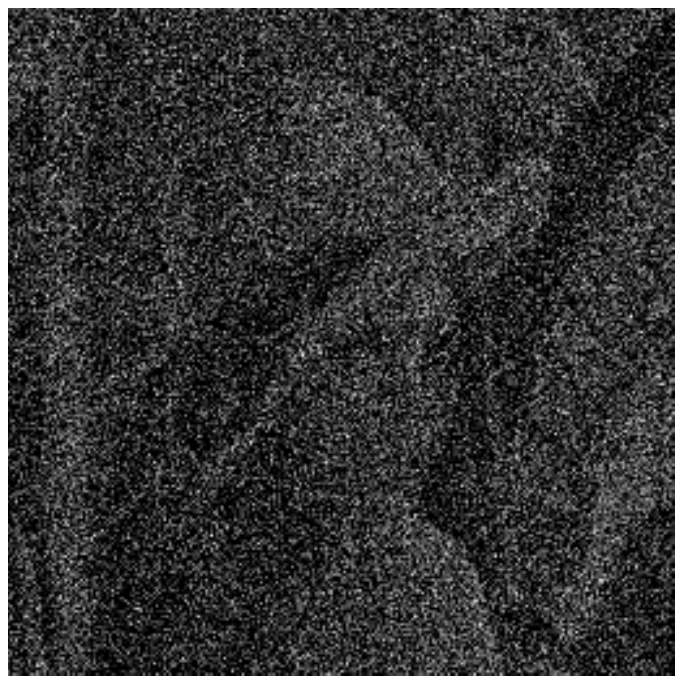}\\
 \vspace{-0.5cm}
  \centerline{(f)}
  \end{minipage}
\caption{Occlusion attack analysis. (a) Encrypted image with $25\%$ occlusion. (b) Decrypted image from (a). (c) Encrypted image with $50\%$ occlusion. (d) Decrypted image from (c). (e) Encrypted image with $75\%$ occlusion. (f) Decrypted image from (e).}
\label{fig8}
\end{figure}

To test the robustness of the proposed encryption algorithm against loss of data, we occlude $25\%$, $50\%$, and $75\%$ of the encrypted image pixels. The decryption process is performed on the occluded encrypted image of ``Lena" with all correct private keys. The encrypted images with $25\%$, $50\%$, and $75\%$ data losses are shown in Fig. \ref{fig8}(a), (c), and (e) while the corresponding recovered images are illustrated in Fig. \ref{fig8}(b), (d), and (f), respectively.  The decrypted images is recognisable even when the encrypted image is occluded up to  $75\%$, though the quality of recovered images drops with the increase in occluded area. We employ the peak signal--to--noise ratio (PSNR) to compute the quality of the recovered
image after attack. For a grayscale image, the PSNR may be computed using the following mathematical expression,
\begin{eqnarray}
\mbox{PSNR}=10~\log_{10}\left(\frac{255^2}{\mbox{MSE}}\right)
\label{eq10}
\end{eqnarray}
where MSE is defined in Eq. (\ref{eq8}). The results of resisting occlusion attack of the proposed algorithm compared to others are presented in Table \ref{Tab4}.

\begin{table}[h]
\caption{Comparison between the PSNR of the reconstructed images with different block size  data loss.}
\label{Tab4}
\begin{center}
\begin{tabular} {| c | c | c | c | }
\hline
    &\multicolumn{2}{c}\bfseries PSNR (dB)&\\
Occlusion (\%)& \multicolumn{2}{c}\bfseries& \\  \cline{2-4}
     & Proposed & Ref. \cite{Lang2012} & Ref. \cite{Rawat2016} \\
\hline

   $25$ & $25.70$ & $13.88$ & $15.42$\\
   \hline
   $50$ & $24.82$ & $11.27$ & $11.26$\\
   \hline
   $75$ & $24.33$ & $9.60$ & $7.05$\\
\hline
\end{tabular}
\end{center}
\end{table}

\subsection{Noise Attack Analysis}
\begin{figure}[h]
\begin{minipage}[b]{0.5\linewidth}
%\vspace{-0.5cm}
  \centering
 \includegraphics[trim= 7cm 10cm -4cm 10cm, clip, width=11cm]{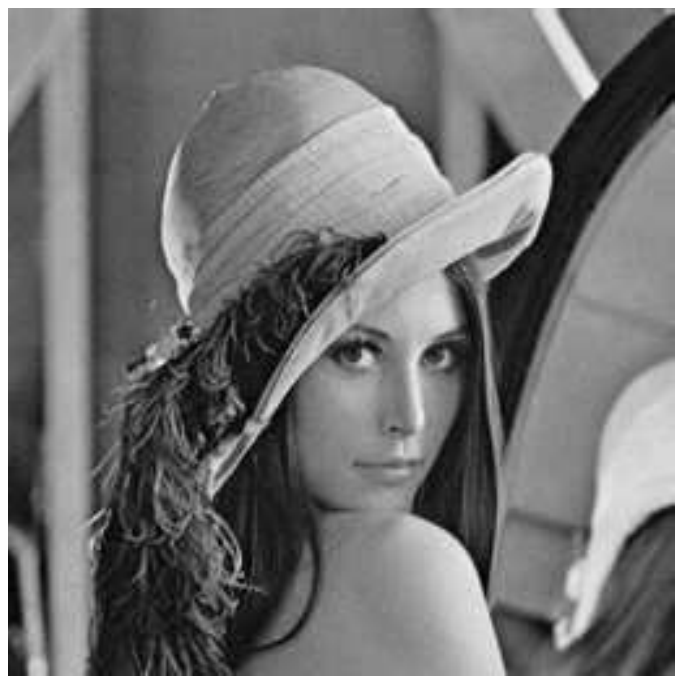}\\
 \vspace{-0.5cm}
  \centerline{(a)}
  \end{minipage}%
\begin{minipage}[b]{0.5\linewidth}
%\vspace{-0.5cm}
  \centering
\includegraphics[trim= 7cm 10cm -4cm 10cm, clip, width=11cm]{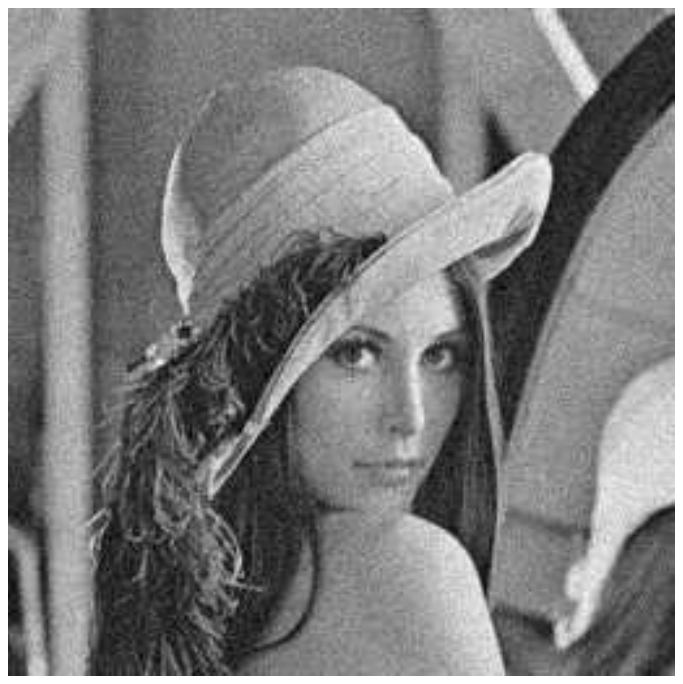}\\
 \vspace{-0.5cm}
  \centerline{(b)}
  \end{minipage}
  \begin{minipage}[b]{0.5\linewidth}
  %\vspace{0.cm}
  \centering
  \includegraphics[trim= 7cm 10cm -4cm 10cm, clip, width=11cm]{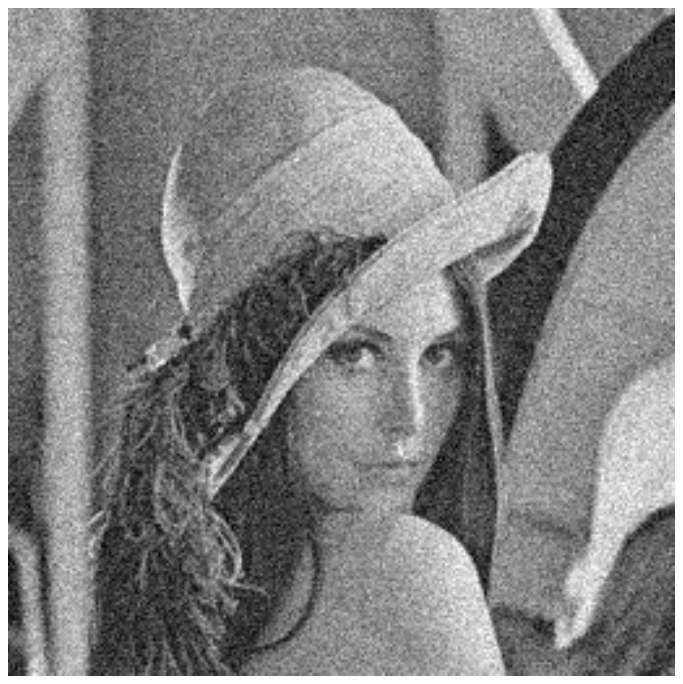}\\
  \vspace{-0.5cm}
  \centerline{(c)}
 \end{minipage}%
\begin{minipage}[b]{0.5\linewidth}
%\vspace{-0.5cm}
  \centering
\includegraphics[trim= 7cm 10cm -4cm 10cm, clip, width=11cm]{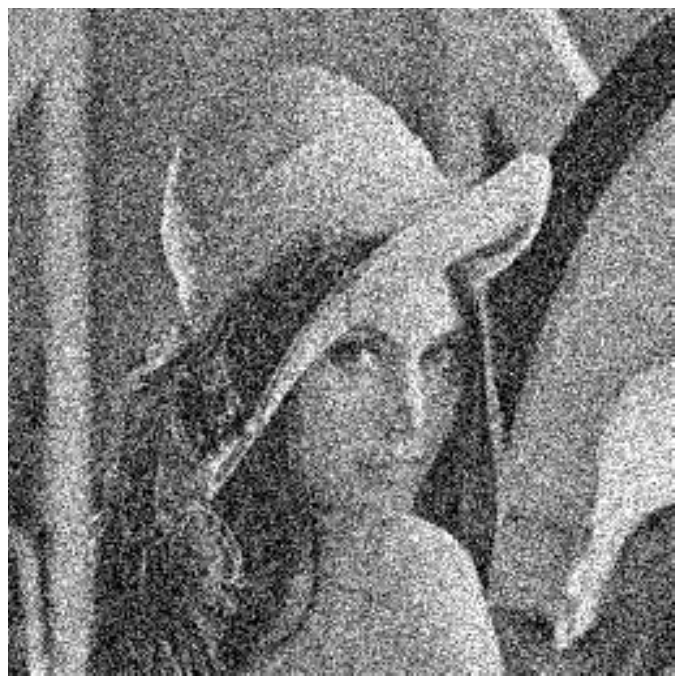}\\
 \vspace{-0.5cm}
  \centerline{(d)}
  \end{minipage}
\caption{Decrypted images with different levels of Gaussian noise. (a) $\sigma=0.01$. (b) $\sigma=0.05$. (c) $\sigma=0.1$. (d) $\sigma=0.2$.}
\label{fig9}
\end{figure}

The robustness of the proposed encryption algorithm is also tested against noise attack. The encrypted Lena image is added to a zero mean Gaussian noise with different intensities. Fig. \ref{fig9}(a)--(d) show decrypted images with noise intensity $0.01$, $0.05$, $0.1$, and $0.2$, respectively. The PSNR values of the proposed scheme and the schemes in \cite{Zhou2016} and \cite{Xiao2017} are compared in Table \ref{Tab5}, which shows that the proposed scheme has better performance at high levels of Gaussian noise. Thus, the results demonstrate that the proposed scheme can resist the noise attack.

\begin{table}[h]
\caption{ The PSNR of decrypted images under different noise levels.}
\label{Tab5}
\begin{center}
\begin{tabular} {| c | c | c | c | }
\hline
    &\multicolumn{2}{c}\bfseries PSNR (dB)&\\
Intensity (\%)& \multicolumn{2}{c}\bfseries& \\  \cline{2-4}
     & Proposed & Ref. \cite{Zhou2016} & Ref. \cite{Xiao2017} \\
\hline

   $0.01$ & $43.97$ & $30.59$ & $38.93$\\
   \hline
   $0.05$ & $32.36$ & $28.55$ & $33.64$\\
   \hline
   $0.1$ & $29.42$ & $25.33$ & $29.57$\\
   \hline
   $0.2$ & $27.95$ & $20.58$ & $26.95$\\
   \hline
   $0.25$ & $27.52$ & $18.91$ & $23.88$\\
   \hline
\end{tabular}
\end{center}
\end{table}

\section{Conclusion}
In summary, an image encryption scheme has been proposed based on logistic maps and 2D discrete Linear chirp transform.  It is demonstrated that the proposed scheme ensures, feasibility, security, and robustness by performing simulations for grayscale images. The results show that the scheme is infeasible to the brute--force attack and offers a great degree of security as seen from its statistical analysis and sensitivity to the encryption parameters. Finally, it has been illustrated that the proposed method is robust against noise and occlusion attacks.

\ifCLASSOPTIONcaptionsoff
  \newpage
\fi

\IEEEtriggeratref{34}


%\begin{biography}[{\includegraphics[width=1in,height=1.25in,clip,keepaspectratio]{mshell}}]{Michael Shell}
%\end{IEEEbiography}
%% insert where needed to balance the two columns on the last page with
%%\newpage


%\vfill


\end{document}